\documentclass{ws-procs975x65}            

\begin{document}

\title{Geons as wormholes of modified gravity}
\author{Gonzalo J. Olmo}

\address{Departamento de F\'{i}sica Te\'{o}rica and IFIC, Centro Mixto Universidad de
Valencia - CSIC. Universidad de Valencia, \\
Burjassot-46100, Valencia, Spain\\
Departamento de F\'isica, Universidade Federal da
Para\'\i ba,\\
58051-900 Jo\~ao Pessoa, Para\'\i ba, Brazil\\
E-mail: gonzalo.olmo@uv.es}
\author{Diego Rubiera-Garcia}
\address{Instituto de Astrof\'isica e Ci\^encias do Espa\c{c}o, Universidade de Lisboa, \\
Faculdade de Ci\^encias, Campo Grande, PT1749-016 Lisboa, Portugal\\
E-mail: drgarcia@fc.ul.pt}

\begin{abstract}
Wormholes may arise as solutions of extensions of General Relativity without violation of the energy conditions. Working in a Palatini approach we consider classical geometries supporting such wormholes. It is shown that the resulting space-times represent explicit realizations of the concept of geon introduced by Wheeler, interpreted as self-consistent bodies generated by an electromagnetic field without sources.
\end{abstract}


\bodymatter

\section{Introduction}

For a long time wormholes have been regarded as exotic solutions of General Relativity (GR), more suitable for science fiction than representing true situations happening in Nature. However, a number of developments and findings in the last few decades, including the seminal paper by Morris and Thorne\cite{MT}, the supernova data\cite{data} suggesting the potential existence of exotic forms of energy driving the accelerated expansion of the universe, and the different approaches to a quantum theory of gravity where topologically non-trivial structures could play a relevant role, have put these once bizarre objects under a new light. Following Visser\cite{Visser}, a (traversable) time-independent, spherically symmetric wormhole space-time can be generically written as

\begin{equation}
ds^2=-e^{2\phi(x)}dt^2 +dx^2 + r^2(x)(d\theta^2+\sin^2 \theta d\varphi^2)
\end{equation}
where $x$ is the proper time. Wormholes are characterized by a number of properties, of which we underline the following:

\begin{itemize}
  \item The coordinate $x$ covers the whole space-time $(-\infty,+\infty)$.
  \item The asymptotic flatness of the two regions connected by the wormhole requires the limits $\lim_{x\rightarrow \pm \infty} \phi(x)=\phi_{\pm}$ to be both finite.
  \item At the asymptotically flat regions, $x \rightarrow \pm \infty$, one has $\lim_{x \rightarrow \pm \infty} r(x)=x$.
  \item The throat of the wormhole satisfies $r_0=\min\{r(x)\}$.
\end{itemize}
Loosely speaking, wormhole are hypothetical tunnels connecting two asymptotically flat portions of the same universe, or two asymptotically flat universes. Historically, the first example was given by the Einstein-Rosen bridge\cite{ER}, which was later shown to be just two copies of the exterior region of a Schwarzschild space-time joined at their event horizons. But with the new insights given by Morris and Thorne in their celebrated 1988 paper \cite{MT} the interest on this field boosted. It is worth pointing out that, within the context of GR, wormholes violate all the pointwise energy conditions, and face the problematic issue of topology change. Because of this, in the past they have been largely regarded as mere theoretical tools for the understanding and teaching of GR.

The related concept of geon - gravitational electromagnetic entities - was introduced by J. A. Wheeler\cite{Wheeler} as hypothetical objects mirroring the idea of body within gravitational physics. Wheeler's original proposal consisted in \emph{balls of light}, an electric beam with so high an intensity that would be held together by its own self-interaction. With the seasoning of non-trivial topology, Misner and Wheeler\cite{MW} were able to give an interpretation of both charge and mass as properties resulting from lines of electric flux trapped in the non-trivial topology of a wormhole. In their picture, geons would represent self-gravitating objects resulting from the Einstein-Maxwell equations without sources with the ambitious goal of explaining all particle properties in terms of non-trivial topologies and fields. The geon program failed, partially due to the lack of explicit, physically motivated, and analytically tractable models.

Starting from a slightly different perspective, in a series of papers\cite{or1,or2,or3,or4,or5,lor13,BId} we have implemented a systematic analysis of classical effective geometries supported by modified theories of gravity. As opposed to the standard procedure in the literature, where a wormhole space-time is given \emph{a priori} and then the Einstein equations are driven back to find the matter sources generating such a geometry, in our approach we derive them from gravitational actions including additional contractions of the Ricci tensor with the metric, and assuming independent metric and affine structures (Palatini approach). This is in sharp contrast with the more standard metric approach, where the connection is given a priori by the Christoffel symbols of the metric (see e.g.\cite{review-mod} for a review on modified gravity). In the last few years we have studied in detail such gravitational actions with electromagnetic fields and found that the black hole point-like singularity of GR is generically replaced by a wormhole structure. Because of their properties, the resulting objects represent explicit implementations of Wheeler's geon within the context of modified gravity.

\section{Wormholes in Palatini gravity}

In Palatini gravity, the field equations admit a GR-like representation of the form

\begin{equation} \label{eq:fieldeqs}
{R_\mu}^{\nu}(q)= \frac{1}{\sqrt{\det \hat{\Sigma}}} \left(\frac{f}{2} {\delta_\mu}^{\nu} + \kappa^2 {T_\mu}^{\nu} \right)
\end{equation}
This representation is valid for $f(R)$ theories \cite{or1}, $f(R,R_{\mu\nu}R^{\mu\nu})$ theories \cite{or2}, Born-Infeld gravity \cite{or3} and in higher-dimensional\cite{BId} and braneworld scenarios\cite{BW}. The matrix ${\Sigma_\mu}^{\nu}$ represents the transformation between the \emph{effective} metric $q_{\mu\nu}$ and the physical metric $g_{\mu\nu}$ as $q_{\mu\nu}= {\Sigma_\mu}^{\alpha} g_{\alpha \nu}$ and, though is model-dependent, can be shown to depend only on the matter stress-energy tensor ${T_\mu}^{\nu}$, and the same applies to the gravity function $f$. The independent connection $\Gamma_{\mu\nu}^{\lambda}$ is compatible with the metric $q_{\mu\nu}$, namely, $\nabla_{\mu} (\sqrt{-q} q^{\alpha\beta})=0$ (but not with $g_{\mu\nu}$, $\nabla_{\mu} (\sqrt{-g} g^{\alpha\beta}) \neq 0$), so it is given by the Christoffel symbols of $q_{\mu\nu}$. The field equations (\ref{eq:fieldeqs}) thus represent a system of second-order field equations, where all the terms on the right-hand-side only depend on the matter. As $g_{\mu\nu}$ is algebraically related to $q_{\mu\nu}$ via the matter sources, the field equations for $g_{\mu\nu}$ are second-order as well. In vacuum, ${T_\mu}^{\nu}=0$, the equations (\ref{eq:fieldeqs}) yield those of GR, which implies the absence of ghost-like degrees of freedom.

In static, spherically symmetric space-times, we take the matter sector to be that of an electromagnetic field. By solving the field equations in different gravitational backgrounds [see Refs.\cite{or1,or2,or3,or4,or5,lor13,BId} for full details] one finds a line element that can be written under the generic (Eddington-Filkenstein) form

\begin{equation} \label{eq:line element}
ds^2=\frac{1}{\Omega_{+}(z)} \left(1-\frac{1+\delta_1 G(z)}{\delta_2 z \Omega_{-}(z)^{1/2}} \right) dv^2 + 2 \frac{dv dx}{\Omega_{+}} + r^2(x)d\Omega^2
\end{equation}
where $z=r/r_c$ is a re-scaled radial coordinate through $r_c$, which typically contains the charge $q$ and some length scale $l_{\epsilon}^2$ encoding the deviations with respect to GR. The constants $\delta_1$ and $\delta_2$ parameterize the solutions in terms of mass, charge and length scale $l_{\epsilon}^2$. The explicit form of the objects $\Omega_{\pm}(z)$ depends on the particular theory of gravity chosen, and the matter-dependent function $G(z)$ typically recovers the GR behaviour, $G(z)\simeq -1/z$, at large distances (provided an asymptotic Coulombian behaviour), but undergoes drastic modifications around $z \simeq 1$.

\begin{figure}[h]
\begin{center}
\includegraphics[width=7.5cm,height=5.5cm]{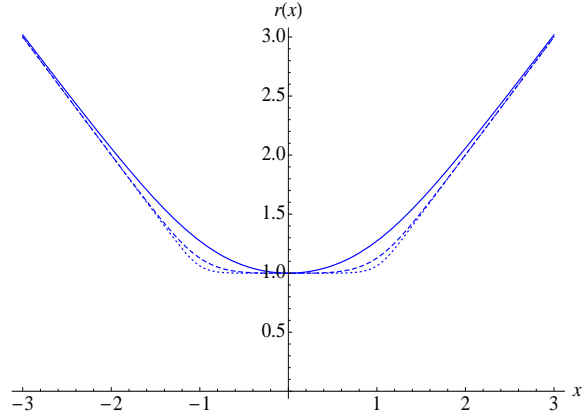}
\caption{Behaviour of the radial function $r(x)$ for Born-Infeld gravity coupled to an electromagnetic field in the case $l_{\epsilon}^2<0$ (see Ref.\cite{or5} for details) in $D=4$ (solid), $6$ (dashed) and $10$ (dotted) space-time dimensions, as given by the expression $r^2(x)=(|x|^{D-2}+(|x|^{2(D-2)}+4r_c^{2(D-2)})^{1/2})/2$, where $r_c=\sqrt{\kappa q l_{\epsilon}}$ with $\kappa=8\pi G$ and $q$ the electric charge. For $r \gg 1$ one has $r^2 \sim x^2$ and the standard GR behaviour. A bounce occurs at $r=r_c$, setting the location of the wormhole throat. \label{fig:1}}
\end{center}
\end{figure}

The space-times (\ref{eq:line element}) above satisfy a number of properties:

\begin{itemize}
    \item Asymptotic flatness is obtained provided that the matter fields satisfy usual energy conditions. Slight modifications of (\ref{eq:line element}) also allow for asymptotically (Anti-)de Sitter solution \cite{or5b}.
    \item For large distances, $r(x) \simeq x$ and the role of $x$ as the standard radial coordinate of the Reissner-Nordstr\"om space-time is restored.
    \item The radial coordinate $r(x)$ reaches a minimum $r=r_c$ at $x=0$ and bounces off [see Fig.\ref{fig:1}]. On this region, large departures from the GR behaviour are found.
    \item The existence or not of horizons depends on the combination of parameters $\delta_1$ and $\delta_2$, namely, on the charge-to-mass ratio.
\end{itemize}
This is in agreement with Visser's requirements introduced discussed above and, therefore, the space-times (\ref{eq:line element}) represent a generalization of the Reissner-Nordstr\"om solution where the point-like singularity is replaced by a finite-size wormhole structure, with the surface $r=r_c$ representing its throat.

In Wheeler's approach\cite{Wheeler,MW} geons are self-gravitating structures where the non-trivial topology of the geon allows to generate both its charge and mass without sources. In our case:

\begin{itemize}

\item The non-trivial topology of the wormhole allows to define the electric charge as the flux of electric lines through a $S^2$ surface enclosing the wormhole throat:
\begin{equation}
\int_{S^2} *F=\pm4\pi q
\end{equation}
where $*F$ is Hodge dual of the electromagnetic field and the sign $\pm$ comes from which side of the wormhole this computation is done. Note that no point-like sources are needed, which is consistent with the sourceless electromagnetic field of the matter sector. A local observer on one of the sides of the wormhole would measure a positive (or negative) charge, though no charges are present in the system and the net global flux is zero, which represents a explicit implementation of the \emph{charge-without-charge} mechanism.

\item Evaluation of the total action (gravitational + electromagnetic) for these geonic solutions produce the generic result $S_T=2M_0 \delta_1/\delta_c c^2 \int dt$ [with $\delta_c$ some constant], where $M_0$ is the total mass of space-time (as given by the mass seen by an asymptotic observer) and the factor $2$ comes from the need of integrating on both sides of the wormhole. This is just the action of a point particle of mass $2M_0\delta_1/\delta_c$. The new gravitational effects are essential for this result, which can be seen as an implementation of the \emph{mass-without-mass} mechanism.

\end{itemize}

In order for these wormholes to be traversable one first requires the absence of horizons which, according to the discussion above, depends on the values of $\delta_1$ and $\delta_2$ for each model. Moreover, one must guarantee its safeness, namely, that a physical observer crossing the wormhole throat is not destroyed on its transit. However, in the wormhole geometries above curvature divergences generically arise at the throat, $r=r_c$, though in some scenarios they can be completely removed for a given mass-to-charge ratio\cite{or2}. We point out that both the existence of the bounce in the wormhole radial function, and some physical properties associated to the geon (like the energy density), are insensitive to the existence or not of curvature divergences [these are generically much milder than their GR counterparts]. Indeed, a three-fold strategy - geodesic completeness, congruence of geodesics, and scattering of waves off the wormhole - has revealed that, in the case of quadratic and Born-Infeld gravity in four dimensions, curvature divergences seem to have little impact (if any) on physical observers, who find a geodesically complete space-time no matter the behaviour of curvature invariants and where no loss of causality occurs among the constituents making up the observer\cite{or7}. One thus concludes that physical observers are not affected by any absolutely destructive effect as they cross the wormhole throat and thus these space-times constitute explicit examples where curvature divergences do not entail space-time singularities.

Let us point out that, as opposed to what happens in the GR case, the fact that we are using a standard electromagnetic field means that the energy conditions are satisfied. The generation of the wormhole structure is a genuine non-perturbative gravitational effect, since only as one gets close to the center of the solutions the wormhole modification of the point-like singularity of GR becomes manifest. The wormhole structure is robust, in the sense that it arises in different gravitational backgrounds and coupled to several kinds of matter, but disappears when the theories are formulated in the standard metric approach. In addition, these wormhole geometries might be generated in dynamical scenarios sourced by high-intensity fluxes of particles carrying mass and charge\cite{or8} or even by large magnetic fields in the early universe\cite{or10}, which could shed new light on the issues of topology change and the geometry of entanglement\cite{or9}.

\section{Conclusions}

In summary, in Palatini theories of gravity, which are supported by the physics of crystalline structures with defects \cite{lor}, self-gravitating, particle-like, non-singular solutions of sourceless equations generated by an electromagnetic field can be found. These classical effective geometries are able to generate a wormhole structure without any need of violation of the energy conditions, and without resorting to the standard thin-shell formalism or \emph{engineering} constructions. Implications of such geonic solutions regarding our understanding of particles and fields are still to be seen. Since we have dealt with a simplified scenario with spherical symmetry and an electromagnetic field it is thus important to investigate if other particle properties (like color charges or spin) can be reproduced by adding other free gauge fields. In this way, geons might potentially yield an interesting phenomenology for gravitational and high-energy physics.

\section*{Acknowledgments}

Work supported by the (Spanish) projects: Ramon y Cajal, FIS2014-57387-C3-1-P, FIS2011-29813-C02-02, i-LINK0780, i-COOPB20105, and the Consolider Program CPANPHY-1205388; the (Portuguese) FCT grants No.~SFRH/BPD/102958/2014 and UID/FIS/04434/2013; and the (Brazilian) CNPq project No.301137/2014-5.

\end{document}